# Identifying Functional Brain Networks of Spatiotemporal Wide-Field Calcium Imaging Data via a Long Short-Term Memory Autoencoder

Xiaohui Zhang[a], Eric C Landsness[b], Lindsey M Brier[c], Wei Chen[g], Michelle J. Tang[b], Hanyang Miao[b], Jin-Moo Lee[b,c,d], Mark A. Anastasio[a*], Joseph P. Culver[c,d,e,f*]

[a]Department of Bioengineering, University of Illinois Urbana-Champaign, Urbana, IL 61801, USA.

[b]Department of Neurology, Washington University School of Medicine, St. Louis, MO 63110, USA.

[c]Department of Radiology, Washington University School of Medicine, St. Louis, MO 63110, USA.

[d]Department of Biomedical Engineering, Washington University School of Engineering, St. Louis 63130, MO, USA.

[e]Department of Electrical and Systems Engineering, Washington University School of Engineering, St. Louis, MO 63130, USA.

[f]Department of Physics, Washington University School of Arts and Science, St. Louis, MO 63130, USA.

[g]Solomon H. Snyder Department of Neuroscience, Johns Hopkins University School of Medicine, Baltimore, MD 21205, USA

* Joseph P. Culver: culverj@wustl.edu, Mark A. Anastasio: maa@illinois.edu

## Abstract

Wide-field calcium imaging (WFCI) that records neural calcium dynamics allows for identification of functional brain networks (FBNs) in mice that express genetically encoded calcium indicators. Estimating FBNs from WFCI data is commonly achieved by use of seed-based correlation (SBC) analysis and independent component analysis (ICA). These two methods are conceptually distinct and each possesses limitations. Recent success of unsupervised representation learning in neuroimage analysis motivates the investigation of such methods to identify FBNs. In this work, a novel approach referred as LSTM-AER, is proposed in which a long short-term memory (LSTM) autoencoder (AE) is employed to learn spatial-temporal latent embeddings from WFCI data, followed by an ordinary least square regression (R) to estimate FBNs. The goal of this study is to elucidate and illustrate, qualitatively and quantitatively, the FBNs identified by use of the LSTM-AER method and compare them to those from traditional SBC and ICA. It was observed that spatial FBN maps produced from LSTM-AER resembled those derived by SBC and ICA while better accounting for intra-subject variation, data from a single hemisphere, shorter epoch lengths and tunable number of latent components. The results demonstrate the potential of unsupervised deep learning-based approaches to identifying and mapping FBNs.

**Introduction**

The brain is a complex network consisting of spatially distributed, but functionally linked, regions that share information with each other (Fox et al. 2005; van den Heuvel and Hulshoff Pol 2010; Power et al. 2011). With advances in neuroimaging across multiple spatial-temporal scales (Irani et al. 2007; Lake et al. 2019; Wheelock et al. 2019), the task of identifying functional brain networks (FBNs) from neuroimaging data remains an important topic of research within the cognitive neuroscience community, with many large scale multi-site studies underway (Mueller, Weiner, Thal, Petersen, C. Jack, et al. 2005; Mueller, Weiner, Thal, Petersen, C.R. Jack, et al. 2005; Fox and Raichle 2007; Power et al. 2011; for the Alzheimer's Disease Neuroimaging Initiative et al. 2017; Elam et al. 2021). Unsupervised deep learning has recently demonstrated encouraging results for the task of decomposing functional magnetic resonance imaging (fMRI) data to identify FBNs in humans (Li et al. 2021 Mar 23). While these advances have provided new capabilities for researchers to study human brain function, their translation to mouse models and novel imaging modalities remains to be investigated. Different from indirect measures of neural activity through a blood oxygen level-dependent (BOLD) signals, wide-field calcium imaging with genetically encoded calcium indicators (GECIs) enables recording of regional neuronal depolarization in mice across the entire cortex. These calcium indicators can be targeted to genetically specified cell types (Ma et al. 2016) and allow for higher spatial-temporal resolution measurement of neural activity (Dana et al. 2014). The development and application of WFCI has stimulated interest in applying various techniques to examine the organization principles in the mouse brain (Wright et al. 2017; Brier et al. 2019; Cramer et al. 2019; Brier et al. 2021; Brier et al. 2022 Aug 16; West et al. 2022).

Over the last decade, a variety functional brain network analyses have been carried out to determine the regional temporal synchrony of blood-based surrogates of neural activity in human brain. Such analyses have been predominantly based on functional magnetic resonance imaging (fMRI) (Fox and Raichle 2007), but also high-density diffuse optical tomography (Eggebrecht et al. 2014; Culver et al. 2016; Wheelock et al. 2019; Uchitel et al. 2022) and functional near infrared spectroscopy (Zhang et al. 2016; Zhang and Zhu 2019; Duan et al. 2020; Ren et al. 2022). The recent development and application of WFCI in small rodents has stimulated interest in applying similar methodologies to identify FBNs in mice (White et al. 2011; Wright et al. 2017; Brier et al. 2019; Cramer et al. 2019; Brier et al. 2021; Brier et al. 2022 Aug 16; West et al. 2022). The seed-based correlation (SBC) (Fox et al. 2005; Wright et al. 2017; Brier et al. 2019) and independent component analysis (ICA) methods (van de Ven et al. 2004; Cramer et al. 2019; West et al. 2021) are two of the most widely applied techniques for identifying functional brain networks from neuroimaging data. SBC permits the identification of brain regions that are functioning together by correlating the time course of a pre-defined seed location and the remaining pixels/voxels time course in the field-of-view (FOV). However, SBC computes bivariate measures and therefore may not comprehensively capture the complex relations underlying brain activity (Salvador et al. 2020; Brier et al. 2021). Alternatively, there are a number of multivariate alternatives being used across various neuroimaging modalities (Craddock 2013; Geerligs et al. 2016; Brier et al. 2021). Among them, ICA is one of the commonly applied multivariate methods that jointly models the relationship among multiple pixels/voxels in the FOV to identify FBNs (van de Ven et al. 2004; Cramer et al. 2019; West et al. 2021). As a data-driven method, ICA does not rely on *a priori* spatial or temporal models to identify the hidden spatial-temporal neural sources from the data (Wang and Guo 2019). By assuming that the sources are spatially independent and have non-Gaussian distributions, spatial ICA (sICA) has been applied for many applications to separate neural signal mixtures into spatially independent, functionally connected brain networks (Xu et al. 2013; Cramer et al. 2019). Nevertheless, the assumption of spatial independence of active brain areas may not be strictly valid (Jung et al.). Furthermore, the unmixing process of traditional ICA is linear, while the brain is a highly nonlinear system (Bi et al. 2018). Thus, a linear decomposition may not be the best way to explore the embedded spatial-temporal information from neuroimaging data (Bi et al. 2018).

Recently, deep learning has surged in popularity across a wide range of neuroimaging applications (Wang et al. 2019; Yan et al. 2019; Cui et al. 2020; Ma et al. 2020; Zhao et al. 2020; Dong et al. 2021; Li et al. 2021 Mar 23; Fan et al. 2022; Zhang, Landsness, Chen, et al. 2022; Li and Fan). Most of these applications involve supervised learning that is typically used in decoding studies to relate neuroimages to behavioral or clinical observations (Zhang, Landsness, Chen, et al. 2022; Zhang, Landsness, Culver, et al. 2022; Zhao et al. 2022). However, the scarcity of labeled neuroimaging data and the data imbalance pose challenges to learn robust and generalizable features by use of supervised models (Zhang, Landsness, Chen, et al. 2022). More recently, unsupervised representation learning methods have been developed for disentangling the sources that drive intrinsic brain activity with large-scale fMRI data (Huang et al. 2018; Dong,

Qiang, Lv, Li, Liu, et al. 2020; Dong, Qiang, Lv, Li, Dong, et al. 2020; Zhang et al. 2020; Kim et al. 2021; Li et al. 2021 Mar 23; Qiang et al. 2021; Qiang et al. 2021; Yan et al. 2021; Dong). For use with fMRI data, a deep convolutional autoencoder (CAE) was proposed to model the temporal patterns of FBNs followed by obtaining corresponding spatial patterns using regression (Huang et al. 2018). However, this approach solely focuses on modeling the temporal dynamics of brain networks, neglecting to simultaneously consider the spatial dynamics inherent within these networks. Later, a spatial-temporal convolutional neural network (ST-CNN) that simultaneously extracts both spatial and temporal characteristics to identify human FBNs was proposed (Zhao et al. 2020). Due to the successful application of sequential autoencoders for dynamic biological signals, a deep sparse recurrent autoencoder (DSRAE) was later introduced to simultaneously extract spatial-temporal patterns for identification of connectome-scale brain networks (Li et al. 2021 Mar 23). While these studies were based on hemoglobin-based assays of neuronal signals, their application to WFCI that directly assays calcium dynamics have not yet been explored.

In this work, a long short-term memory (LSTM) autoencoder (AE) is employed to simultaneously learn spatial-temporal latent embeddings from wide-field calcium imaging data for identifying functional brain networks. The aim of this study is to qualitatively and quantitatively assess the ability of the LSTM autoencoder with regression (LSTM-AER) to identify FBNs from spatial-temporal WFCI data of mice. We also compare the performance of the LSTM-AER against conventional SBC and ICA. The impact of several analysis options that include the amount of spatial-temporal data and the pre-defined number of latent components on the identified FBNs is investigated. This study will provide insights and avenues for researchers to effectively employ unsupervised recurrent autoencoder to estimate FBNs from WFCI in broader applications in the future.

**Materials and Methods**

1. Animals and surgical procedures

The study was approved by the Washington University School of Medicine Animal Studies Committee and followed the guidelines of the National Institutes of Health *Guide for the Care and Use of Laboratory Animals*. A total of N=21 transgenic mice (12-16 weeks of age, sex) expressing GCaMP6f in excitatory neurons (driven by *Thy1* promoter) were acquired from Jackson Laboratories (G57BL/6J-Tg (Thy1-GCaMP6f) GP5.5Dkim; stock: 024276). Mice were housed in 12-hour light/dark cycles with lights on at 6:00 AM and given ad lib access to food and water.

Experimental paradigms were similar to previous studies (Zhang, Landsness, Chen, et al. 2022; Landsness et al.) and are briefly summarized here. Before the data acquisition, the head of mouse was shaven, and a midline incision was made to expose the skull. Two stainless steel EEG self-tapping screws were fixed at approximately -1 mm posterior to bregma, and +/- 4.5 mm lateral to bregma with one reference EEG screw in the cerebellum. To record muscle activity, a 203 micrometer Teflon coated EMG wire (A-M Systems, Sequim, Washington, catalog #792100) was threaded into the neck muscle and referenced to the cerebellum.

2. Wide-field calcium imaging

Mice were acclimated to head fixation while secured in a black felt hammock for one to three sessions ranging to 30-180 minutes until the EEG/EMG signals demonstrate the presence of sleep. Each mouse underwent a 3-hour undisturbed WFCI session while being able to move freely with its head secured. During the imaging session, the mice were placed under a cooled, frame-transfer EMCCD camera overhead (iXon 897, Andor Technologies, Belfast, Northern Ireland, UK) and four collimated LEDs, as described previously (Wright et al. 2017; Brier et al. 2019; Zhang, Landsness, Chen, et al. 2022). Along with concurrent EEG recording, sequential illumination was provided by four LEDs: 454 nm (GCaMP6 excitation), 523 nm, 595 nm, and 640 nm at a frame rate of 16.8 Hz per channel. The field of view was approximately 1 $cm^2$ with pixel resolution of 78 $\mu m^2$ which covers the dorsal surface of the brain.

3. Image processing

WFCI recordings were processed with a custom MATLAB package (Brier and Culver 2021) as follows. A binary mask that isolates the brain region was manually annotated and the time traces of all pixels were spatially and temporally detrended. The modified Beer-Lambert Law was applied on reflectance changes in the 523 nm, 595 nm, and 640 nm LED channels to solve for relative fluctuation of the oxygenated-hemoglobin (HbO$_2$) and deoxygenated-hemoglobin (HbR). The GCaMP fluorescence signal was corrected for absorption by HbO$_2$ and HbR using a ratiometric approach where the reflectance channel at the GCaMP6 emission wavelength (523 nm) was considered as a reference. Specifically, the final corrected GCaMP6 time traces for a given pixel was defined as

$$y(t) = \frac{I^{em}(t)}{I^{ref}(t)} \cdot \frac{I_0^{ref}}{I_0^{em}},$$

where $I^{em}$ refers to the detected fluorescence emission intensity at time point $t$, and $I^{ref}$ denotes the measured reflectance changes at the emission wavelength. Images were smoothed with a $5 \times 5$ Gaussian filter. The global signal averaged across all brain pixels was regressed out. Image sequences were affine-transformed to a common Paxinos (Paxinos and Franklin 2019) space using the positions of bregma and lambda.

Recordings of each mouse were divided into 5-min continuous sessions followed by reshaping each session into a two-dimensional (2D) data matrix $\mathbf{X} \in \mathbb{R}^{T \times N}$, where $T$ is the number of frames and $N$ is the number of brain pixels. These data matrices will be used as inputs for both ICA and the LSTM-AER. Three sessions from each mouse were held-out to evaluate the methods in this study.

4. Seed-based correlation method

Seed locations in the motor, somatosensory, retrosplenial, visual and auditory cortices were chosen based on the Paxinos atlas (Paxinos and Franklin 2019). A 0.5 mm diameter circle at each seed location was averaged to create a seed time trace. Pearson correlation was computed between the seed time traces and every other brain pixel. Correlation coefficients were Fisher Z-transformed and standardized to zero mean and unit variance. Functional connectivity (FC) matrices were computed by taking the Pearson correlation of the average time traces between two seed regions. The Z-transformed correlations were average across all subjects and inversed Z-transformed to obtain group-average FC matrix.

5. Independent component analysis (ICA)

ICA is an approach that aims to recover a set of maximally independent sources given their observed multivariate linear mixture without knowledge of the source signals or the mixing parameters (Comon 1994). Spatial ICA has been widely applied on spatial-temporal neuroimaging data to extract independent components (ICs) that represents functional brain networks (Mckeown et al. 1998; Calhoun and Adalı 2012; Beheshtian et al. 2021 Jul 20; Weiser et al. 2021; West et al. 2021). Given an input data matrix $\mathbf{X} \in \mathbb{R}^{T \times N}$, where $T$ is the number of time points (frames) and $N$ denotes the number of pixels in the brain region (two-dimensional data are reshaped into one-dimension), the spatial ICA model can be described as:

$$\mathbf{X} = \mathbf{AS},$$

Where $\mathbf{S} \in \mathbb{R}^{C \times N}$ is the source matrix consisting of $C$ spatially independent components and $\mathbf{A} \in \mathbb{R}^{T \times C}$ is a mixing matrix consisting of time traces of the corresponding ICs. One of the popular approaches for implementing ICA is FastICA algorithm (Hyvärinen and Oja 2000). FastICA iteratively estimates an unmixing matrix $\mathbf{W} = \mathbf{A}^{-1}$, $\mathbf{W} \in \mathbb{R}^{C \times T}$, such that $\hat{\mathbf{S}} = \mathbf{W}^T \mathbf{X}$ and each $\mathbf{s} = (s_1, s_2, \ldots, s_N)^T$ is mutually independent. This is achieved by maximizing the non-Gaussianity of the estimated independent components. In this study, $C = 16$ was empirically chosen as the default setting, guided by previous studies (Makino et al. 2017; West et al. 2021). The pre-defined number of components was further varied to investigate its effect on the identification of FBNs. The estimated spatial components were converted to z-score values for visualization. FastICA was implemented using scikit-learn package (Pedregosa et al. 2011). For each subject (N=21), three data matrices constructed from 5-min WFCI recordings were used to evaluate our model. By default, ICA in this study is referred to single-session spatial ICA (Choe et al. 2015) for consistency in comparison to other methods used in this study.

6. Identifying functional brain networks using LSTM-AER

6.1 Network architecture and training details

A previous study has demonstrated the feasibility of simultaneous spatial-temporal decomposition of fMRI data by use of an unsupervised recurrent autoencoder(Li et al. 2021 Mar 23). Different from fMRI, WFCI enables simultaneous capturing of neural dynamics with higher temporal resolution and signal-to-noise ratio and can achieve cell type specificity. In this study, as part of the LSTM-AER method, we used an LSTM autoencoder, operating on wide-field calcium imaging data from mice, to learn spatial-temporal embeddings for identification of functional brain networks.

A recurrent neural network is a type of network with feedback loops that uses internal memory to process sequential data. Among them, the LSTM cells - consisting of memory units controlled by a gating mechanism - allow for maintaining long-term memory. This architecture can permit feature learning from WFCI data that possess high temporal resolution and relatively long scan times. It is worth noting that the response time of GECI's is around ~100ms, compared to ~4s of fMRI BOLD. Thus, GECI is a 40x faster neural reporter compared to BOLD. A 5 min GECI scan has the equivalent number of independent frames as a 200 min BOLD scan. The architecture of the LSTM autoencoder used in our study and an illustration of LSTM unit are shown in Figure 1a and Figure 1b, respectively. The autoencoder is a 10-layer deep neural network composed of an encoder and a decoder. Given a spatially flattened 2D input data matrix $\mathbf{X} \in \mathbb{R}^{T \times N}$, the encoder network maps $\mathbf{X}$ into a low-dimensional latent embedding $\mathbf{Y} \in \mathbb{R}^{T \times C}$ via a fully connected layer to reduce the spatial dimensionality, followed by three LSTM layers to capture the dependency along the temporal dimension. The decoder networks with symmetric architecture mirroring that of the encoder subsequently reconstruct $\hat{\mathbf{X}}$ from the 2D spatial-temporal latent embeddings $\mathbf{Y}$. By default, the three LSTM layers in the encoder consist of 64, 32, and 16 units, and the three LSTM layers in the decoder include 32, 64 and 64 units, respectively. Each of the fully connected layers and LSTM layers were activated with the tanh function. The number of LSTM units in the bottleneck layer, which pre-defines the number of spatial components to decompose, was varied to investigate its effect on FBN identification.

The LSTM autoencoder was trained and validated on 549 and 69 data matrices, respectively, that were formed from 5-min continuous segments of WFCI recordings. A test set consisting of three data matrices from each of the individual mice (N=21) were held-out to evaluate the model. The autoencoder was trained to minimize the mean square error between the input data matrix $\mathbf{X}$ and the reconstructed data matrix $\hat{\mathbf{X}}$ by use of an Adam optimizer (stochastic gradient descent) with a learning rate of 0.001 for 200 epochs. The model with the best validation performance was selected. The autoencoder was implemented in Python 3 with TensorFlow 2.2.0 using NVIDIA GPUs.

6.2 Ordinary least square regression

Given a data matrix $\mathbf{X}$ and the corresponding encoded latent embedding $\mathbf{Y}$, the corresponding spatial maps $\mathbf{W}$ representing FBNs were derived by ordinary least square regression as:

$$\mathbf{W} = \underset{\mathbf{W} \in \mathbb{R}^{C \times N}}{\mathrm{argmin}} \|\mathbf{X} - \mathbf{YW}\|^2,$$

where C is the number of spatially independent components and N is the number of brain pixels. Each row of **W** can be reshaped into a 2D spatial map by use of the brain mask. The regression procedure is shown in Figure 1c. All spatial components were standardized into z-score values for visualization. For simplicity, the hybrid pipeline described in this Section 6 will be denoted as LSTM-AER in the rest of the paper.

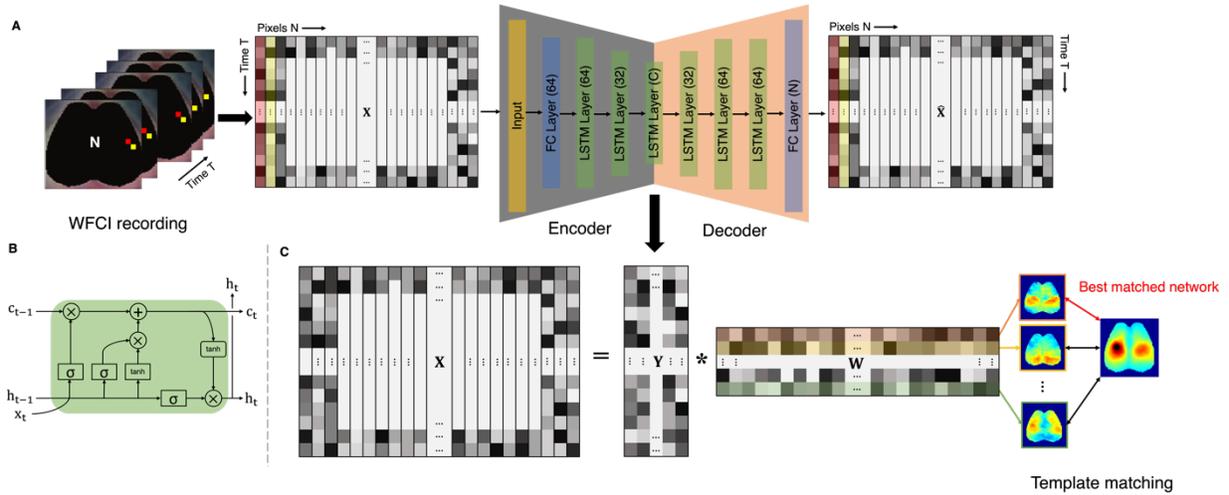

Figure 1 Schematic of the proposed LSTM-AER approach for identifying functional brain networks from spatial-temporal wide-field calcium imaging data. (A) The architecture of the LSTM autoencoder with the number of units of each layer listed in the parenthesis. **X** is the input data matrix constructed from 5-min WFCI recording and $\hat{\mathbf{X}}$ is the reconstructed data matrix from the output of the autoencoder. C denotes the pre-defined number of latent components and $C = 16$ by default. (B) An illustration of the LSTM memory unit with gates to control the content of inputs $x$, cell states $c$ and outputs $h$. (C) The ordinary least square regression used to estimate the spatial maps of FBNs from the extracted latent embeddings **Y** and original input data matrix **X**. A template matching procedure was followed to identify the best matched brain network.

7. Template matching

To identify the brain networks in the derived in spatial maps **W**, derived by LSTM-AER, a template matching procedure was used as described in previous studies (Garrity et al. 2007; Greicius et al. 2007; Zuo et al. 2010; Rocca et al. 2014; Bi et al. 2018; Mejia et al. 2020 Jun 4; Tahedl and Schwarzbach 2020; Wu et al. 2020). This technique is similar to typical identification methods used with ICA. The group-level SBC maps averaged across subjects (N=21) generated from all available 5-min WFCI sessions described in Section 4 were considered as (ground truth) templates. Functional brain networks including motor, somatosensory, retrosplenial, visual and auditory cortices were considered as networks of interest. Spatial correlation coefficients were computed between the estimated spatial maps and the templates. The FBN identify of each estimated spatial map was assigned based on the maximum correlation coefficient value.

For both the ICA and LSTM-AER methods, functional network connectivity (FNC), which is considered as a higher level of FC, was computed as pairwise Pearson correlation between the time traces corresponding to the networks of interest (Jafri et al. 2008; Joel et al. 2011; Arbabshirani et al. 2013; Choe et al. 2015; Du et al. 2021; Salman). The values were Fisher Z-transformed, averaged across all subjects and then inverse Z-transformed to obtain group-average FC matrices. Conceptually, FC matrices from SBC are different from those computed by use of ICA and LSTM-AER (Joel et al. 2011). In this study, we only aim to compute FC/FNC matrices for all three methods to provide a qualitative and quantitative comparison.

8. Evaluation

8.1 Comparison to axonal projection connectivity maps

Axonal projection connectivity (APC) maps were collected from the Allen Mouse Brain Connectivity (AMBC) Atlas via http://connectivity.brain-map.org (Lein et al. 2007; Oh et al. 2014). These maps reflect enhanced green fluorescent protein (EGFP)-labeled axonal projection of sites injected with pan-neuronal adeno-associated virus under the promoter *Syn1* in C57BL/6J mice. APC images were selected based on the proximity of the injection sites to those used in this study of WFCI data. APC maps were converted into projection images using the AMBC cortical map signal viewer. A brain mask was applied, and APC maps were affine-transformed into the same atlas space as our WFCI data. Each APC map was normalized by its maximum fluorescence intensity prior to analysis. Following the previous schemes(Bauer et al. 2018; Brier et al. 2021), APC maps were thresholded at 50% intensity while spatial maps of FBNs derived from SBC, ICA and LSTM-AER were thresholded at a median of positive z-score. All thresholded maps were binarized and the dice coefficient was calculated between the APC maps and thresholded spatial maps derived from each technique described above. Paired t-tests were performed to assess the statistical difference.

8.2 Subject variation

To visualize and compare the subject variation of the FBNs estimated from LSTM-AER, ICA, and SBC, t-distributed Stochastic Neighbor Embedding (t-SNE) (Maaten and Hinton 2008) was applied to visualize the spatial layout of the FBNs in a two-dimensional space. The t-SNE is a technique that projects high-dimensional data points into a set of embedded points in a low-dimensional space whose similarities mimic those of original points. The Silhouette index was computed to measure the similarity of FBNs within the same subject compared to those of other subjects.

8.3 Inter-subject reproducibly and intra-subject reliability

To measure the inter-subject reproducibility and intra-subject reliability of FBNs estimated by use of the LSTM-AER, ICA and SBC methods, spatial correlation was computed for pairwise subject-level spatial maps between subjects and pairwise single-session spatial maps within subjects, respectively. A network-wise paired t-test was performed on the inter- and intra-subject reproducibility to assess the statistical difference.

8.4 Stability of FBNs from various epoch lengths

Since GCaMP6 can probe neural activity at higher frequencies than traditional hemodynamic-based functional connectivity, it is of interest to examine the effect of epoch length on the stability of the FBNs. Spatial correlation coefficients were computed to measure the spatial similarity between the FBN maps derived from various epoch length (5s up to 300s) and group-averaged maps created by use of WFCI data with a conventional 5-min duration.

9. Data and code availability

To promote further applications and analyses by external groups, a subset of the WFCI data described in this study are publicly available on PhysioNet (Goldberger et al. 2000; Landsness et al.). The code for data pre-processing can be accessed via https://github.com/brierl/Mouse_WOI (Brier and Culver 2021). All network training and testing code are available at https://github.com/comp-imaging-sci/brain-networks-autoencoder.

**Results**

1. Spatial and temporal patterns of FBNs identified via LSTM-AER

Our first goal was to qualitatively evaluate the spatial and temporal patterns of FBNs identified by use of the proposed LSTM-AER, as compared with the traditional spatial ICA and SBC method. The group-averaged spatial maps over 21 mice are displayed in Figure 2A. For classification of spatial maps estimated by the ICA and LSTM-AER methods, as described above, we calculated the spatial similarity of each map with the FC templates and assigned them to network that corresponds to highest score. The spatial maps produced by use of the LSTM-AER were found to be similar to those from traditional ICA and SBC, while less focalized patterns were observed (Figure 2C). When compared with fluorescence images representing APC, spatial maps of FBNs from LSTM-AER proved to overlay as well as those acquired by the other methods. However, the pixel-wise inter- and intra-subject standard deviation maps (Supplemental Figure 1) demonstrated that the spatial maps corresponding to the LSTM-AER were more consistent across mice and runs, compared with the ones estimated from ICA and SBC. We evaluated the spatial similarity of among FBNs derived by the LSTM-AER, ICA and SBC methods (Supplemental Figure 2), and confirmed that the networks estimated by the three methods were similar but not identical.

Beyond the spatial distribution of the estimated FBNs, it is also of interest to understand the temporal synchronization of the identified networks. When assessing the functional connectivity for the three methods, FNC matrices were obtained by computing Pearson's correlation coefficients between the time traces of networks derived by the LSTM-AER and ICA methods, and FC matrices were computed using the pairwise time traces extracted from seeds (Figure 2D). The functional network connectivity matrices lacked the broadly dispersed positive and negative connection values that are seen with ICA and SBC, supported by the histogram (Figure 2E) of matrix values from Figure 2D. The largely positive values for matrix of LSTM-AER resided along the main diagonal and the homotopic contralateral off diagonal. The functional connectivity matrices corresponding to the LSTM-AER method demonstrated better network specificity and left-right symmetric Compared to ICA and SBC. Also, the LSTM-AER method demonstrated better network specificity in off-diagonal networks and left-right symmetric.

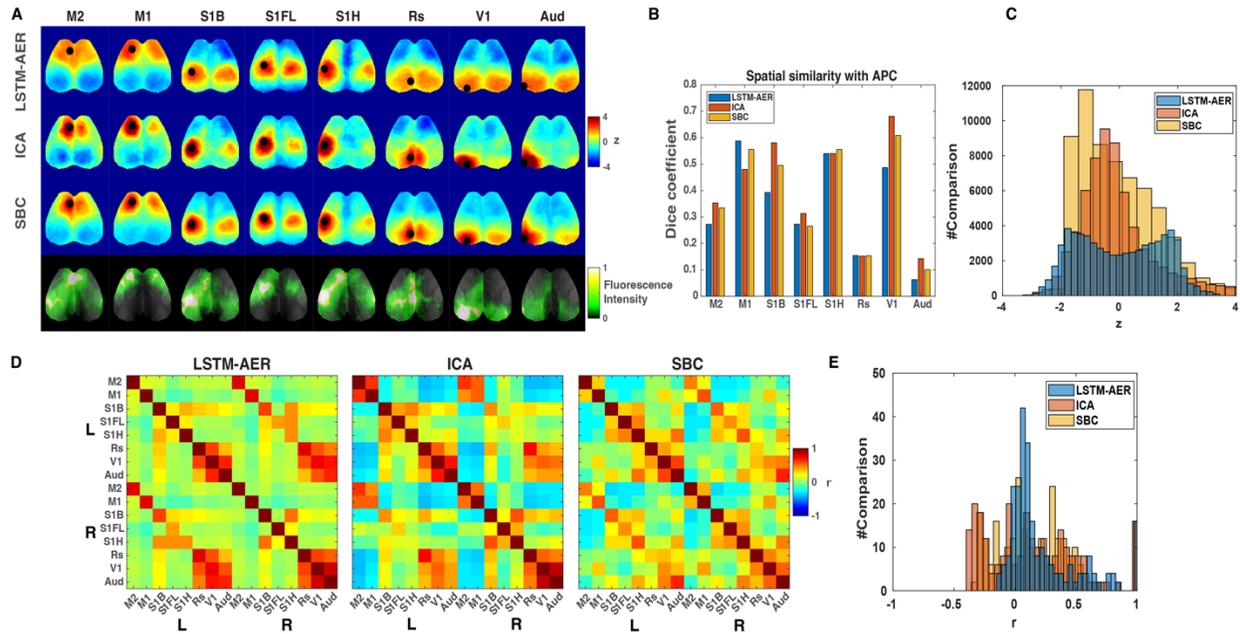

Figure 2 Spatial and temporal patterns of FBNs identified by use of the LSTM-AER, spatial ICA and SBC method. (A) Group-average (N=21) spatial maps of FBNs identified by use of LSTM-AER along with the ones derived from ICA and SBC. The black circle denotes the seed location used in SBC. The bottom row shows the APC maps with injection sites in proximity to the corresponding FBNs. (B) Dice coefficient between the APC maps and the average spatial maps of FBNs estimated from LSTM-AER, ICA, and SBC, respectively. (C) Histogram plotting the number of comparisons for all z-values of spatial maps in (A). (D) Average (N=21) network-based connectivity matrices for each method. (E) Histogram plotting the number of comparisons from (C) for each correlation value.

2. Individual subject variation

Whereas the analyses described above focus on the group-level spatial maps of the FBNs, we are also interested in how the distribution of the identified FBNs vary across individual mice. Therefore, the t-SNE method was employed to visualize individual high-dimensional spatial maps of FBNs (more than 10,000 pixels) in a low-dimensional manifold (Figure 3A). Points that are close in high-dimensional space would remain close in the new, low-dimensional space. Reducing the dimensionality of the data is critical for many applications, as it allows avoiding redundancy, compact visualization and identify latent features in the data (Casanova et al. 2021). The clusters of FBNs from individual mice were apparent in the 2D representation of FBNs obtained from LSTM-AER and SBC, but not those from ICA. This distinction was further quantified by computing the Silhouette index to measure the degree of clustering by each mouse (Figure 3B). Both the Silhouette value for LSTM-AER (mean±std: 0.8128±0.3808) and SBC (0.8131±0.4747) were significantly higher than that of ICA (mean±std: -0.2517±0.1468, paired t-test, $p<0.001$), indicating that the FBNs identified by the proposed LSTM-AER better captured individual subject variation than ICA.

Another aim was to investigate the in intra-subject reliability (independent time points) and inter-subject reproducibility (independent mice) of the FBNs estimated by use of the different methods. For each mouse (N=21) with three runs, the intra-subject reliability was assessed by verifying the spatial correlation between corresponding FBNs from each of the two runs (Figure 3C). The inter-subject reproducibility was operationalized as the spatial correlation between averaged spatial maps that were computed based on each mouse for corresponding FBNs, respectively (Figure 3D). Results indicate that both the intra-subject reliability and the inter-subject reproducibility of FBNs identified by LSTM-AER and SBC are statistically higher than those computed using single-session ICA. In addition, the intra- and inter-subject standard deviations maps (Supplemental Figure 1) also confirmed the increased reliability and reproducibility in the FBNs estimated from LSTM-AER and SBC, while those from single-session ICA are more variant. These observations imply that the LSTM-AER is less susceptible to physiological and acquisition artifacts as compared to ICA, and thus allows for identifying more robust FBNs that are consistent across subjects.

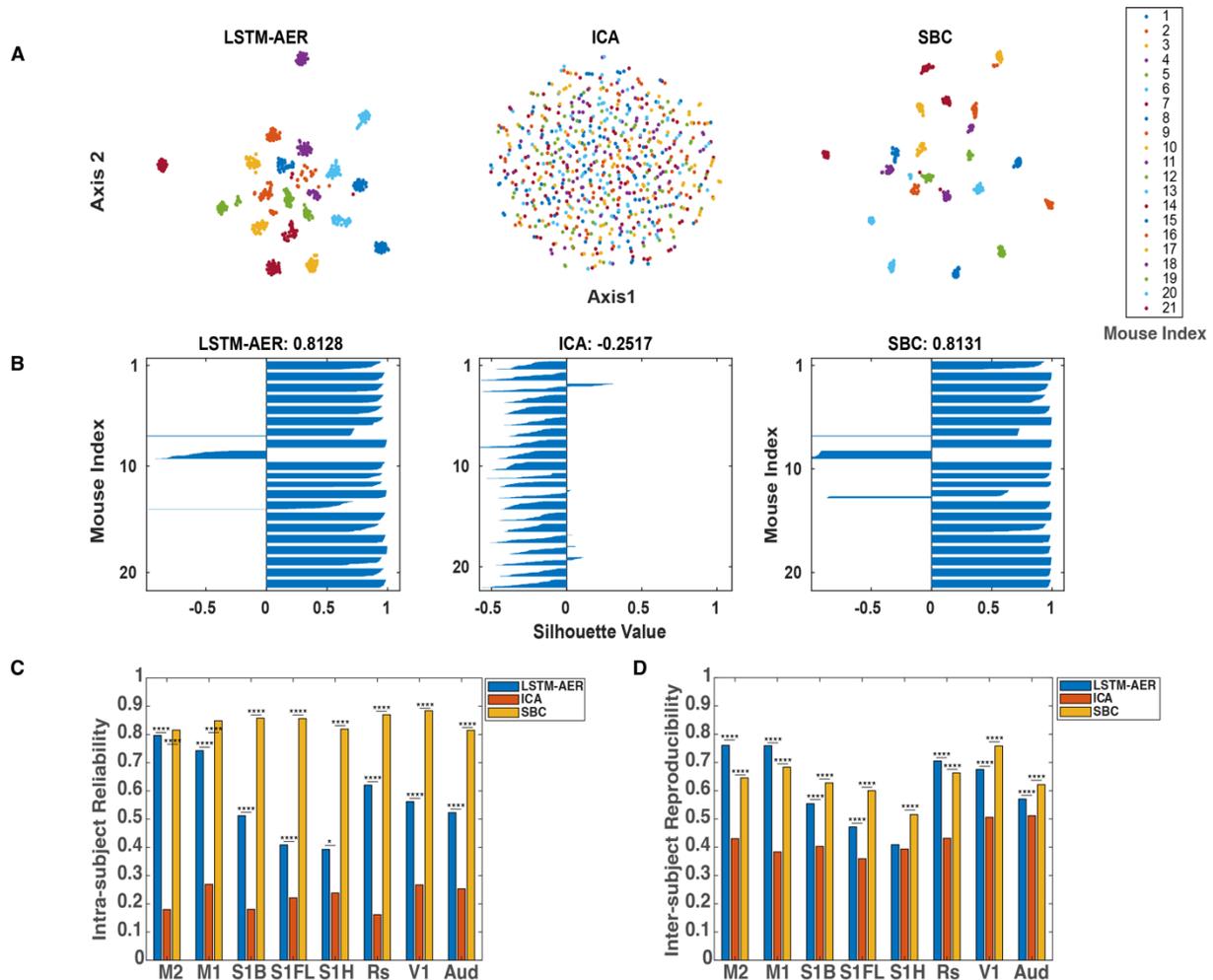

Figure 3 Individual variation of FBNs identified by use of the LSTM-AER, ICA and SBC methods. (A) Spatial maps of mouse-wise FBNs estimated with the LSTM-AER, ICA and SBC were visualized in a 2D-space with t-SNE. Different color codes index different mice. (B) The Silhouette value that measures how well 2D representation of FBNs acquired with t-SNE from different runs matches within one mouse (points labeled with the same color) as the others (points labeled with different colors), compared across the LSTM-AER, ICA and SBC. (C) Intra-subject reliability measured by average spatial similarity of FBNs identified with the LSTM-AER, ICA and SBC within each mouse. *=p<0.05; ****=p<0.001. (D) Inter-subject reproducibility measures average spatial similarity of FBNs identified with LSTM-AER, ICA and SBC between different mice. *=p<0.05; ****=p<0.001.

3. Impact of varying the amount of spatial and temporal data in identification of FBNs

The spatial-temporal nature of WFCI offers the possibility to explore the impact of quantity of spatial and temporal information in the data on the identified FBNs across different approaches. For different epoch lengths ranging from 5s to 300s, FBNs were identified using LSTM-AER, ICA and SBC. Spatial maps of FBNs based on 10s WFCI data are shown in Figure 4A. Different from the maps in Figure 2A where 5-min runs were used to identified the FBNs, the signal-to-noise ratio of the spatial maps with shorter epoch lengths of 10s were lowest among all three methods for comparison. A spatial similarity analysis was employed to evaluate how the estimated FBNs vary with increasing epoch length of WFCI data. The spatial correlation coefficient between each of the spatial maps and the corresponding group-average (N=21) maps calculated from 5-min runs is shown in Figure 4B. In general, the epoch length necessary to estimate FBNs that no longer change significantly was found to be method-dependent. In contrast to ICA where mild improvement of the average spatial similarity compared to group-average reference was observed along with the increasing epoch lengths, the LSTM-AER method achieved a spatial similarity score of 0.70 when an epoch length of 240s was utilized. However, SBC outperformed the LSTM-AER method and ICA at all selected epoch lengths and the spatial similarity reached 0.8 upon convergence based on 180s of data.

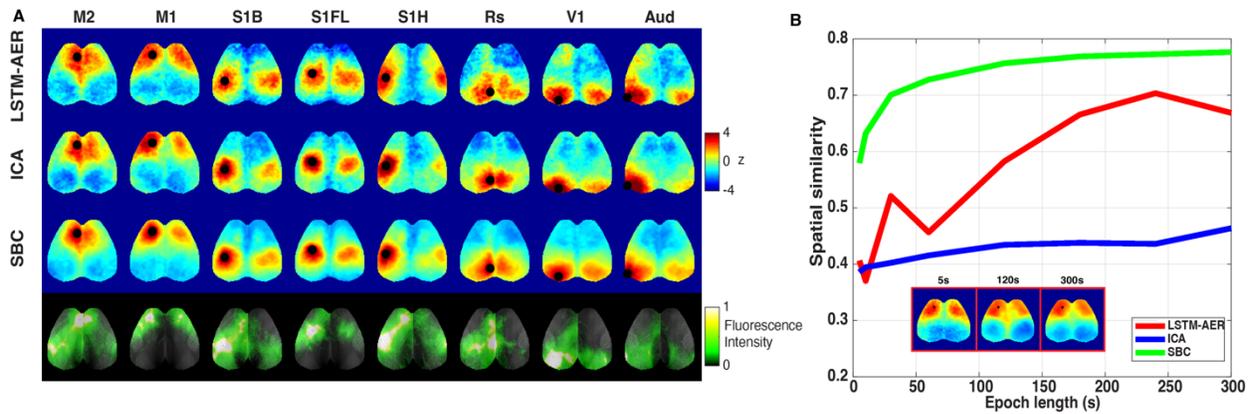

Figure 4 Varying the amount of temporal data to identify FBNs. (A) Group-average (N=21) spatial maps of FBNs identified by use of the LSTM-AER method, ICA and SBC using 10-s epoch lengths. The bottom row shows the APC maps with injection sites in proximity to the corresponding FBNs. (B) Spatial similarity measured by spatial correlation between each of the spatial maps from individual runs of various epoch lengths and the group-average (N=21) maps of 5-min runs, averaged over all networks and all testing runs. The spatial maps of FBNs identified by use of the LSTM-AER method are highlighted as examples.

In addition to varying the epoch length to investigate how the amount of temporal data affects the estimated FBNs, we assessed whether the proposed LSTM-AER remains effective if only a subset of pixels was used for training. To evaluate this, we constrained the data to cover only a single hemisphere (left) and FBNs were identified using the LSTM-AER method, ICA and SBC. When using only hemispheric data, FBNs could still be effectively identified by all three methods (Figure 5A). Compared with APC images, the spatial maps estimated by use of the LSTM-AER method were observed to overlay as well as, or better than, those produced by ICA and SBC (Figure 5B).

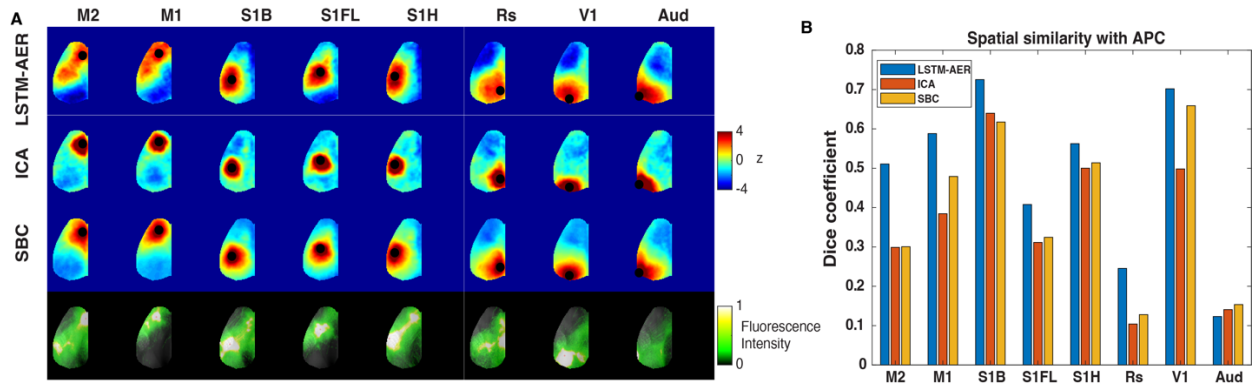

Figure 5 Varying the amount spatial information to identify FBNs. (A) Group-average (N=21) spatial maps of FBNs identified by use of LSTM-AER, ICA and SBC using hemispheric (left) pixels only. The black circle denotes the seed location used in SBC. The bottom row shows the APC maps masked with only left hemisphere. (B) Dice coefficient between the APC maps and the average spatial maps of FBNs estimated from LSTM-AER, ICA, and SBC using hemispheric data only, respectively.

4. Impact of the tunable number of latent sources on the identified FBNs

There is currently no consensus regarding how to determine the number of latent sources employed by independent component analysis (ICA) (Li et al. 2007; Wang and Li 2015). When employing the proposed LSTM-AER method, we also tried to seek more insights into the specification of the latent space dimension in the LSTM autoencoder. Here, the number of latent sources C was varied to investigate is impact on the identified FBNs. A small order of 8 (Figure 6A,B,C,D) and a larger order of 32 (Figure 6E,F,G,H) were selected as both the number of hidden units in the bottleneck layer of the LSTM autoencoder as well as the number of independent components in ICA, respectively. When the number of latent sources was underestimated, such as for C=8, some of the known FBNs were not identified by the LSTM-AER method (e.g., M2 and S1H). For both the LSTM-AER method and ICA, the spatial patterns of FBNs lack focality (Figure 6B) compared to their corresponding counterparts derived with C=16 (Figure 2C). Similarly, for

C=8, the FNC matrices computed by pairwise time traces of FBNs revealed more dispersed patterns due to underestimation of sources from the WFCI data. Conversely, when extracting C=32 sources, the structure of the FBNs yielded by the LSTM-AER method and ICA were finer and more focal (Figure 6F) compared to the corresponding maps of each method in the case of C=16. In contrast to the LSTM-AER method, the maps yielded ICA revealed higher-intensity activation of subnetworks from the larger networks. The FNC matrices effectively captured the covarying patterns between the FBNs pairs, while they were similar as those found at C=16 since consistent network templates were employed throughout the study.

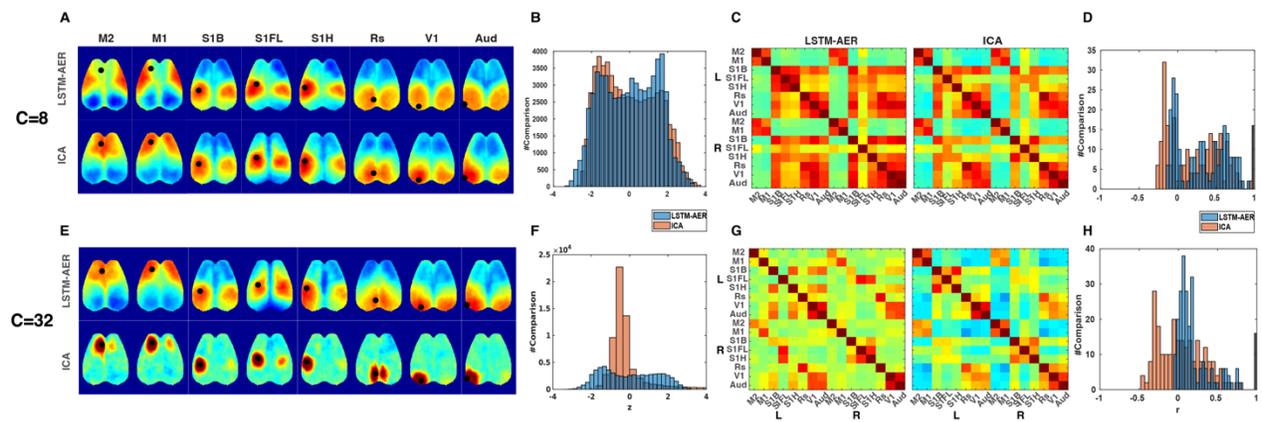

Figure 6 Effect of tuning the number of latent sources (C=8, C=32) on identification of FBNs. (A)(B) Group-average (N=21) spatial maps of FBNs identified by use of LSTM-AER and ICA. (B)(F) Histogram plotting the number of comparisons from (A) and (E) for each z value. (C)(G) Average FNC matrices by computing temporal correlation between pairwise FBNs. (D)(H) Histogram plotting the number of comparisons from (C) and (G) for each correlation value.

## Discussion

In this study, we adapted and extended the use of a LSTM autoencoder to identify FBNs from spatial-temporal wide-field calcium imaging (WFCI) in mice. A hybrid, two-step pipeline LSTM-AER, was proposed that involved an LSTM autoencoder to identify spatial-temporal latent sources from WFCI data, followed by ordinary least square linear regression to estimate the spatial patterns of FBNs. The proposed approach allowed us to effectively learn spatial-temporal latent sources and estimate FBNs from a large-scale WFCI dataset. The ability of the LSTM-AER method to identify FBNs was evaluated through comparisons to other popular functional connectivity methods, specifically, SBC and ICA.

Conceptually, the LSTM autoencoder adapted from (Li et al. 2021 Mar 23) seeks to identify the spatial-temporal latent sources nonlinearly from wide-field calcium data collected from a group of subjects in an unsupervised manner. This is a different approach from SBC, which computes simple bivariate correlations between the seed locations and pixels, and ICA, in which latent sources are linearly decomposed. The spatial and temporal characteristics of the LSTM-AER identified FBNs possessed similarities compared to traditional bivariate SBC method and spatial ICA. However, the LSTM-AER method demonstrated several distinct performance advantages over SBC and ICA including the capability of capturing subject variation, better concordance with axonal tracer maps given limited spatial and temporal information, and allowing for tunable number of latent components to identify FBNs.

1. Spatial and temporal patterns yielded by the LSTM-AER method, ICA and SBC were generally similar, though not identical

While the LSTM-AER method and ICA are conceptually distinct in the way they decompose latent sources from the data, both are categorized as blind source separation techniques in which the underlying "sources" of the data are unknown. The output from the proposed LSTM-AER method and ICA are spatial maps that reveals synchronous brain regions and the time traces corresponding to the identified networks. In contrast, SBC computes Pearson correlation on time series from preselected seed regions and derived cross-correlations spatial maps with pixels in the FOV. The spatial patterns of FBNs produced by the LSTM-AER method, ICA and SBC shares similarities with the axonal tracer maps, but they are not identical. The spatial patterns of FBNs yielded by the LSTM-AER method appeared to less focal compared to those identified by ICA and SBC (Figure 2A, B, C). This could be a result of several factors. First, the use of the L2 loss was only enforced between the input and the output data matrices to promote data consistency while no sparsity constraints were imposed through the network. However, it is known that sources underlying brain activities are sparse (Jääskeläinen et al. 2022). Thus, a variation of the LSTM-AER that encourages sparsity in the bottleneck layer such as *k*-sparse autoencoder (Makhzani and Frey 2014) could be investigated in the future. Second, ordinary least square regression was employed to estimate the weight maps from the learned latent sources of the data. More advanced regression strategies that promote sparsity, such as least absolute shrinkage and selection operator (Lasso) and ElasticNet regression, can also be considered.

When computing functional temporal connectivity, a practice adopted in the LSTM-AER approach and ICA is to compute the correlation among FBNs rather than focusing upon independent pairwise correlation between brain regions in the SBC method. While conceptual differences exist, we observed that the network-based connectivity matrix lacks the disperse positive and negative connections that were seen with ICA and SBC (Figure 2D, E). This indicates that the FBNs identified by the LSTM-AER method were largely unaffected by the negative connectivity pattern potentially introduced by data preprocessing steps such as global signal regression (Murphy et al. 2009; Liu et al. 2017). These results further suggest that the learning-based LSTM-AER method, may be more robust to varying data preprocessing procedures (different versions of global signal regression) across imaging groups.

2. The LSTM-AER method identified reproducible FBNs that also reflected subject variation

When applying SBC analysis and single-session ICA, spatial patterns of FBNs are generated for individual runs of each mouse from the cohort independently. To identify consistent FBNs underlying group neuroimaging data, group ICA has been widely applied to multi-subject human fMRI studies (Chen et al. 2008). A key feature of the LSTM-AER method is that it identifies spatial-temporal latent sources from data collected from a cohort of mice, and from multiple runs of continuous WFCI recordings. Data matrices from different mice at various time points are given as inputs to the autoencoder network. These separate mouse specific data matrices, reflect subject variation. Hence, the FBNs identified from sources extracted by LSTM-AER may also reveal subject variation while finding the consistent FBNs across the cohorts of mice. The subject-wise geometries visualized with t-SNE (Casanova et al. 2021) confirm that the learning-based LSTM-AER approach effectively discriminates FBNs from various mice as compared to SBC and ICA in low-dimensional manifolds (Figure 3A, B), which indicates that the identified FBNs could be further considered as fingerprints to identify an individual mouse for different tasks.

While FBNs projected onto 2D manifolds reflect clustering that align with the subject identity better, the reproducibility of these FBNs within and between each of the subjects has not been previously addressed by brain connectome studies (Zuo et al. 2010; Franco et al. 2013; Soltysik 2020). Both the inter-subject reliability and intra-subject reproducibility are higher with the proposed LSTM-AER method as compared to the other blind source separation method using ICA, where decomposition computed with individual runs fails to capture the common patterns shared by the cohorts (Figure 3C, D). Interestingly, SBC also produces somewhat similar spatial patterns as those from the LSTM-AER, which was not observed with ICA. This could be explained by the fact that SBC is inherently dependent on the priori anatomical or functional seeds information. These analyses of the intra- and inter- subject variability confirmed the robust and stable nature of the intrinsic brain activity recorded by WFCI, and the intrinsic self-organization is functionally similar across individual mice (Franco et al. 2013). It also reminds us of the necessity and importance of assessing the reliability and reproducibility of FBNs when employing novel methods to process WFCI data. Future studies should reassess reproducibility using the seed locations informed by the ICA and LSTM-AER (Franco et al. 2013). Specification of the epoch length that can yield enhanced reproducibility for each of the methods can also be investigated in future studies (Brier et al. 2021).

3. The LSTM-AER method is robust to limitations in spatial or temporal information

Ongoing research efforts are focused on determining the ideal amount of temporal data needed to construct typical and stable FBNs (Korhonen et al. 2021). While an appropriate length of the epoch should allow one to capture the changes in FBNs, too short an epoch length may result in noisy FBN estimates (Figure 4A). Conventionally, an epoch length of 5 min is used to identify and analyze the functional brain network by use of WFCI data (Wright et al. 2017; Brier et al. 2019; Brier et al. 2021). More recently, it has been reported that the FBNs identified by use of SBC were relatively stable with as little as 30-second delta-band (0.4-4 Hz) WFCI data, while slight drift observed with an extended length (Wright et al. 2017). Here, we observed that, while the stability of FBNs estimated by SBC converges similarly as reported previous work when various epoch lengths of the data are considered, FBNs identified by the LSTM-AER method and ICA were drastically different. The spatial similarity of FBNs yielded by the LSTM-AER method to that of group-average data improved with increasing amount of temporal WFCI data, whereas the spatial similarity of ICA demonstrated only mild improvement (Figure 4B). This could be a result of the superior representation learning capabilities of the LSTM-AER method compared to ICA. Nevertheless, a slight drift occurred when longer epochs were considered for both methods, which could be attributed to neurophysiological changes that occur. These observations highlight the importance of selecting an appropriate epoch length for FBN identification, with consideration of different neuroimaging modalities and frequency-bands.

The spatial-temporal nature of WFCI offers the possibility to alter the amount of the spatial data to understand FBNs identified using hemispheric WFCI data. To push the limits of the LSTM-AER method and understand whether it is still capable of identifying common FBNs given limited spatial information, hemispheric WFCI data were given as input for the LSTM-AER method and ICA. Surprisingly, the hemispheric FBNs identified by use of the LSTM-AER method overlaid more precisely compared to those from ICA and most of the FBNs yielded by SBC (Figure 5). In studies involving fMRI, spatial ICA has often been used because the number of voxels is typically much larger than the number of scans (Calhoun et al. 2001; van de Ven et al. 2009). When hemispheric data are provided, the temporal dimension is larger than the spatial dimension since WFCI affords higher temporal resolution relative to hemodynamics (Wright et al. 2017; Brier et al. 2019; Brier et al. 2021), whereas ICA requires a large number of samples to perform well (Smith et al. 2012). In contrast, the LSTM-AER method was still able to identify FBNs effectively under such conditions, and yielded better spatial similarity compared to ICA (Figure 5B). This indicates the simultaneous identification spatial-temporal latent sources by the LSTM-AER method may be more robust than ICA when the amount of spatial or temporal data varies.

4. The tunable pre-defined number of latent sources allows flexibility for investigation of FBNs

One crucial user-defined parameter to set when applying blind source separation methods, including the LSTM-AER approach or ICA, is the selection of the number of latent sources to be decomposed. This choice can significantly affect the resultant FBNs (Li et al. 2007; Lu et al. 2017). The number of latent sources is usually determined empirically and subjectively (Lu et al. 2017; Makino et al. 2017; Cramer et al. 2019; West et al. 2022). Selection of number of pre-defined sources has a significant impact on the spatial organization of resultant FBNs because networks can be fractionated into sub-networks as the number increases (Kiviniemi et al. 2009). In contrast, underestimating number of pre-defined sources can lead to failure to identify true hidden patterns underlying the brain activities. In our study, we found that a low number of pre-defined latent components resulted in degraded performance in the template matching procedure that led to miss-localization of the large-scale FBNs (Figure 6A, B, C, D). On the other hand, a larger model order yielded more focal networks, and this phenomenon was more obvious with ICA as compared to the LSTM-AER method (Figure 6E, F, G, H). This could be because ICA separates data into maximally independent components and tends to extract more spatially focal networks (Bhinge et al. 2019), whereas the FBNs estimated by use of the LSTM-AER method robustly localize the common networks. It is also worth noting that a useful specification of the latent sources is related to the signal-to-noise ratio of the dataset. How to estimate the number of informative components to reduce over/underfitting remains an open question in the neuroscience community (Li et al. 2007). Thus, when applied in different scenarios, it is advisable to tune the pre-defined number of latent sources in the bottleneck of the autoencoder network and assess the variability of identified FBNs to identify the appropriate number of latent sources. Additionally, the availability of the ground-truth templates and robustness of the template matching procedure plays a role in deciding what scale of network to be identified (Tahedl and Schwarzbach 2020). In the future, more advance approaches to classify the identified spatial maps of FBNs can be investigated. For example, the recently proposed deep learning-based network SiameseICA can be trained with limited number of samples while being able to accurately classify ICA-derived components into different resting states networks (Chou et al. 2022).

The translation of functional brain network analysis methods from fMRI to WFCI remains limited. Some interesting topics remain to be explored in future research. For instance, the recurrent autoencoder is essentially a type of deterministic autoencoder and has non-regularized latent space. Recently, generative models such as variational autoencoders (VAEs) that regularize the latent space have been demonstrated to be a valuable addition for disentangling sources from fMRI activity in human studies. This success motivates the application of such tools to WFCI data. When a limited number of WFCI datasets labeled by human professionals are accessible, it may also be interesting to employ a semi-supervised autoencoder in which an autoencoder is optimized towards hybrid image reconstruction and classification task to investigate the effect of the labeled behavioral/consciousness states on the identification of the FBNs.

**Conclusion**

In this study, we investigated the use of an LSTM autoencoder with regression (LSTM-AER) to identify functional brain networks from spatial-temporal wide-field calcium imaging in mice. The unsupervised LSTM-AER approach can effectively identify similar but not identical FBNs as those produced by commonly employed methods such as seed-based correlation and independent component analysis. FBNs estimated by use of the LSTM-AER method reflect subject variation while producing reproducible spatial patterns compared to ICA. The spatial-temporal dynamics of neural activity recorded by WFCI allows for deeper insights into its effectiveness in identifying FBNs across various imaging methods and applications. The tunable number of pre-defined latent sources in blind source separation methods permit flexibility for analyzing FBNs. This study could promote the broader application of deep learning-based unsupervised representation learning for FBN analysis with WFCI.

## Funding

This work was supported in part by the National Institute of Neurological Disorders and Stroke (R01NS099429 to J.P.C., R37NS110699 and R01NS094692 to J.M.L., K08NS109292-01A1 to E.C.L.), National Institute on Aging (F30AG061932 to L.M.B.), American Academy of Sleep Medicine Foundation (201-BS-19 to E.C.L.), American Heart Association (20CDA35310607 to E.C.L.) and National Institute of Biomedical Imaging and Bioengineering (P41EB031772 to M.A.A.)

## Disclosure statement

The authors declare no potential conflicts of interest.

## Author contributions

**Xiaohui Zhang**: Conceptualization, Methodology, Software, Validation, Formal analysis, Investigation, Writing - Original Draft. **Eric C. Landsness**: Resources, Data curation, Writing - Original Draft, Funding acquisition. **Hanyang Miao**: Resources, Data curation. **Wei Chen**: Resources, Data Curation. **Michelle J. Tang**: Resources, Data curation. **Lindsey M. Brier**: Resources, Data curation. **Jin-Moo Lee**: Funding acquisition. **Mark A. Anastasio**: Conceptualization, Supervision, Writing – Review & Editing, Funding acquisition, Project administration. **Joseph P. Culver**: Conceptualization, Supervision, Project administration, Writing – Review & Editing, Funding acquisition.

**Supplemental Figures**

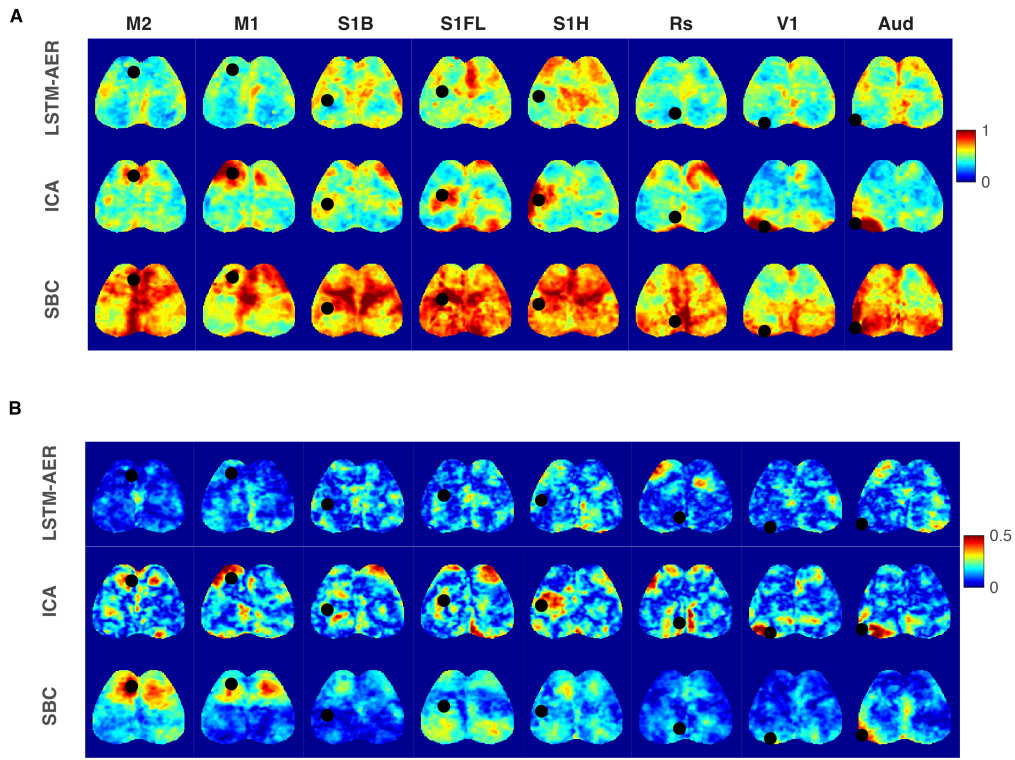

Supplemental Figure 1 Standard deviation maps of FBNs identified by use of the LSTM-AER, spatial ICA and SBC method. (A) Inter-subject standard deviation maps measured on spatial maps of FBNs averaged from each mouse. (B) Intra-subject standard deviation maps measured on spatial maps of FBNs within multiple test sessions of each mouse, averaged over 21 mice.

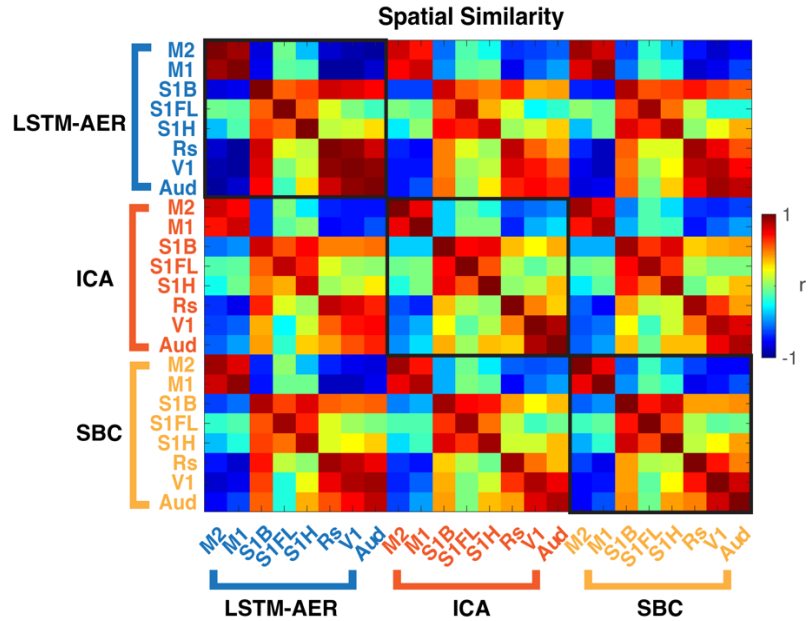

Supplemental Figure 2 Spatial correlation of group-averaged FBNs derived from the LSTM-AER, spatial ICA and SBC method.